\documentclass[11pt]{article}
\usepackage{moriond,graphicx,xspace}

\def\Journal#1#2#3#4{{#1} {\bf #2}, #3 (#4)}

\def\PLB{{\em Phys. Lett.}  B}
\def\PRL{\em Phys. Rev. Lett.}
\def\PRD{{\em Phys. Rev.} D}

\def\be{\begin{equation}}
\def\ee{\end{equation}}
\def\bea{\begin{eqnarray}}
\def\eea{\end{eqnarray}}

\def\Bmumu{\ensuremath{B^0_{(s)} \rightarrow \mu^+\mu^-}\xspace}
\def\Bsmumu{\ensuremath{B^0_s \rightarrow \mu^+\mu^-}\xspace}
\def\Bdmumu{\ensuremath{B^0 \rightarrow \mu^+\mu^-}\xspace}
\def\D0mumu{\ensuremath{D^0 \rightarrow \mu^+\mu^-}\xspace}
\def\Dpimumu{\ensuremath{D^+_{(s)} \rightarrow \pi^+\mu^+\mu^-}\xspace}
\def\Dppimumu{\ensuremath{D^+ \rightarrow \pi^+\mu^+\mu^-}\xspace}
\def\Lbppi{\ensuremath{\Lambda_b^0 \rightarrow p \pi^-}\xspace}
\def\LbpK{\ensuremath{\Lambda_b^0 \rightarrow p K^-}\xspace}
\begin{document}
\vspace*{2cm}
\title{RARE B AND CHARM DECAYS AT THE TEVATRON}

\author{T. KUHR\\
on behalf of the CDF and D0 collaborations}

\address{Institut f\"ur Experimentelle Kernphysik, Universit\"at Karlsruhe (TH),\\
Wolfgang-Gaede-Str. 1, 76131 Karlsruhe, Germany}

\maketitle\abstracts{
The measurements of rare decays are highly sensitive to physics beyond the standard model.
In this article limits on the branching ratios of the decays
\Bmumu, \D0mumu and \Dppimumu are presented.
Furthermore the first measurement of the branching fraction and CP asymmetry of
\Lbppi and \LbpK decays is described.
Data samples with an integrated luminosity of up to 2 fb$^{-1}$ collected
at the Tevatron $p\bar{p}$-collider at $\sqrt{s}=1.96$ TeV were used in these analyses.
The results are consistent with the standard model predictions and 
tighten the constraints on new physics models.}

\section{Introduction}

The branching ratio of a rare decay mode is an interesting quantity to measure
because the contribution from physics beyond the standard model, which can be
negligible compared to the standard model (SM) contribution in the dominant decay modes,
may be sizable in the rare decay mode.
Decays are suppressed in the SM for different reasons.
One of them is that flavor-changing neutral current (FCNC) processes are forbidden at tree level.
The FCNC decays discussed here are \Bmumu, \D0mumu and \Dpimumu.
Although the decays \Lbppi and \LbpK are allowed at tree level in the SM, they are suppressed 
because the $b$ to $u$ quark transition involves the small CKM matrix element $V_{ub}$.

In order to be able to observe rare heavy flavor decays it is essential to produce a sufficient
amount of bottom and charm hadrons.
The large $b\bar{b}$ and $c\bar{c}$ cross section at the Tevatron, more than four orders of magnitude 
higher than at the B-factories, allows to probe very small branching ratios.
Another advantage of the Tevatron is the production of all species of $b$ hadrons
so that rare decays of $B^0_s$ mesons and $b$ baryons can be studied.

On the other hand the inelastic cross section is $10^3$ times higher than $\sigma(b\bar{b})$
requiring very selective and efficient triggers.
For rare decays in particular triggers on pairs of muons or pairs of displaced tracks are used.
Another challenge at the Tevatron is the high combinatorial background from fragmentation tracks.
Sophisticated selection procedures based on kinematic, topologic and particle identification
quantities are employed to extract the signal.

\section{Rare decay measurements}

\subsection{\Bmumu}

The FCNC process \Bsmumu is predicted to have a branching ratio
of $\mathcal{B}(\Bsmumu)=(3.42 \pm 0.54)\times 10^{-9}$ in the SM\cite{Buras:2003td}.
The \Bdmumu decay is further suppressed compared to the $B^0_s$ decay by $|V_{td}/V_{ts}|^2$.
The SM prediction for the branching ratio is $\mathcal{B}(\Bdmumu)=(1.00 \pm 0.14)\times 10^{-10}$.
A significant enhancement of the $B^0_s$ and $B^0$ branching ratios is predicted by
several new physics models.
For example in the minimal super-symmetric standard model (MSSM) the $B^0_s$ branching ratio 
is proportional to $\tan^6\beta$ where $\tan\beta$ is the ratio between the vacuum expectation values
of the two neutral Higgs fields.
In $R$-parity violating super-symmetric (SUSY) models an enhancement is possible even at low 
values of $\tan\beta$.

\begin{figure}[h]
\begin{center}
\parbox{0.45\textwidth}{
\includegraphics[width=0.45\textwidth]{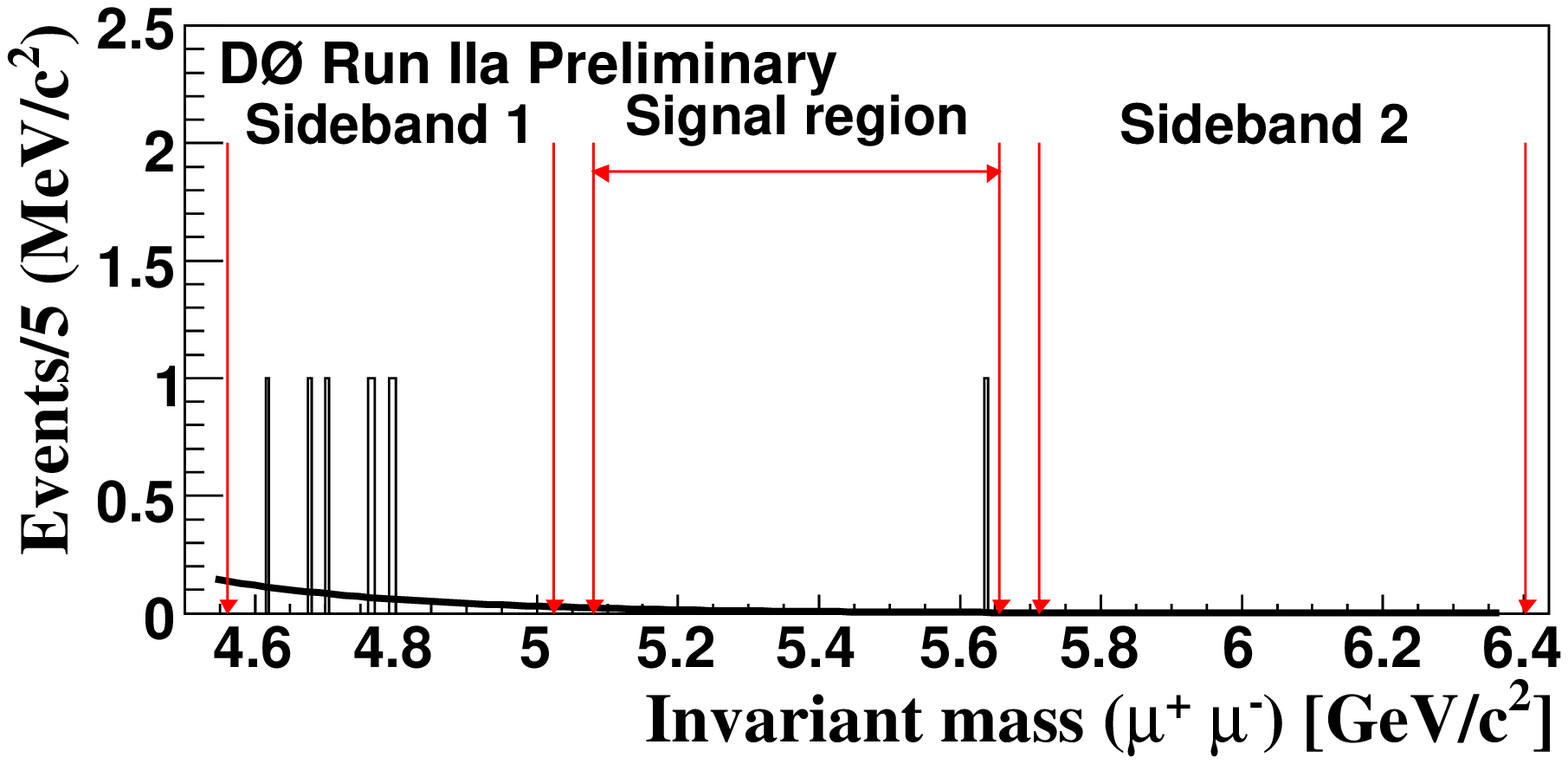}
\includegraphics[width=0.45\textwidth]{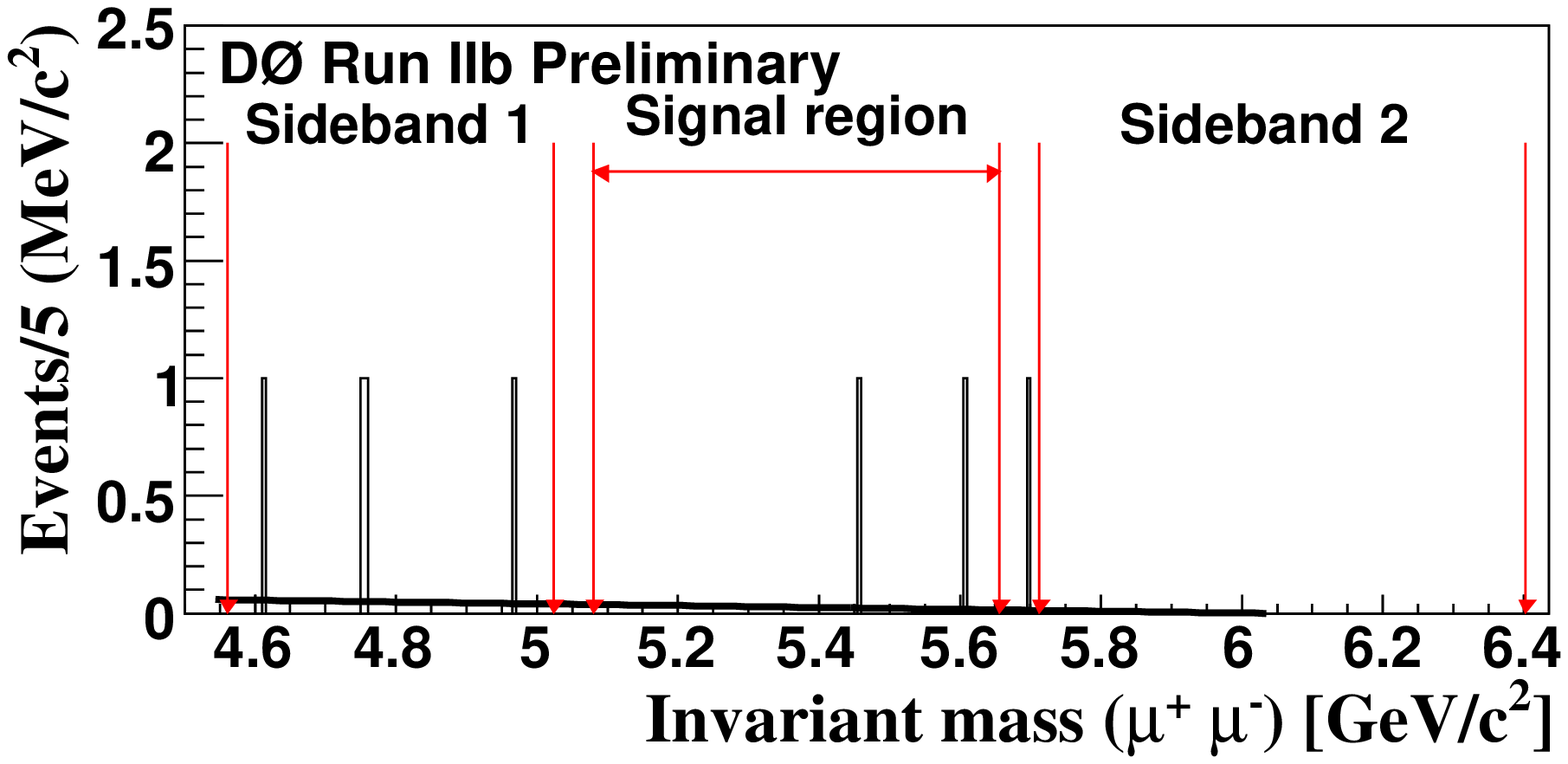}
}
\parbox{0.09\textwidth}{
}
\parbox{0.45\textwidth}{
\includegraphics[width=0.4\textwidth]{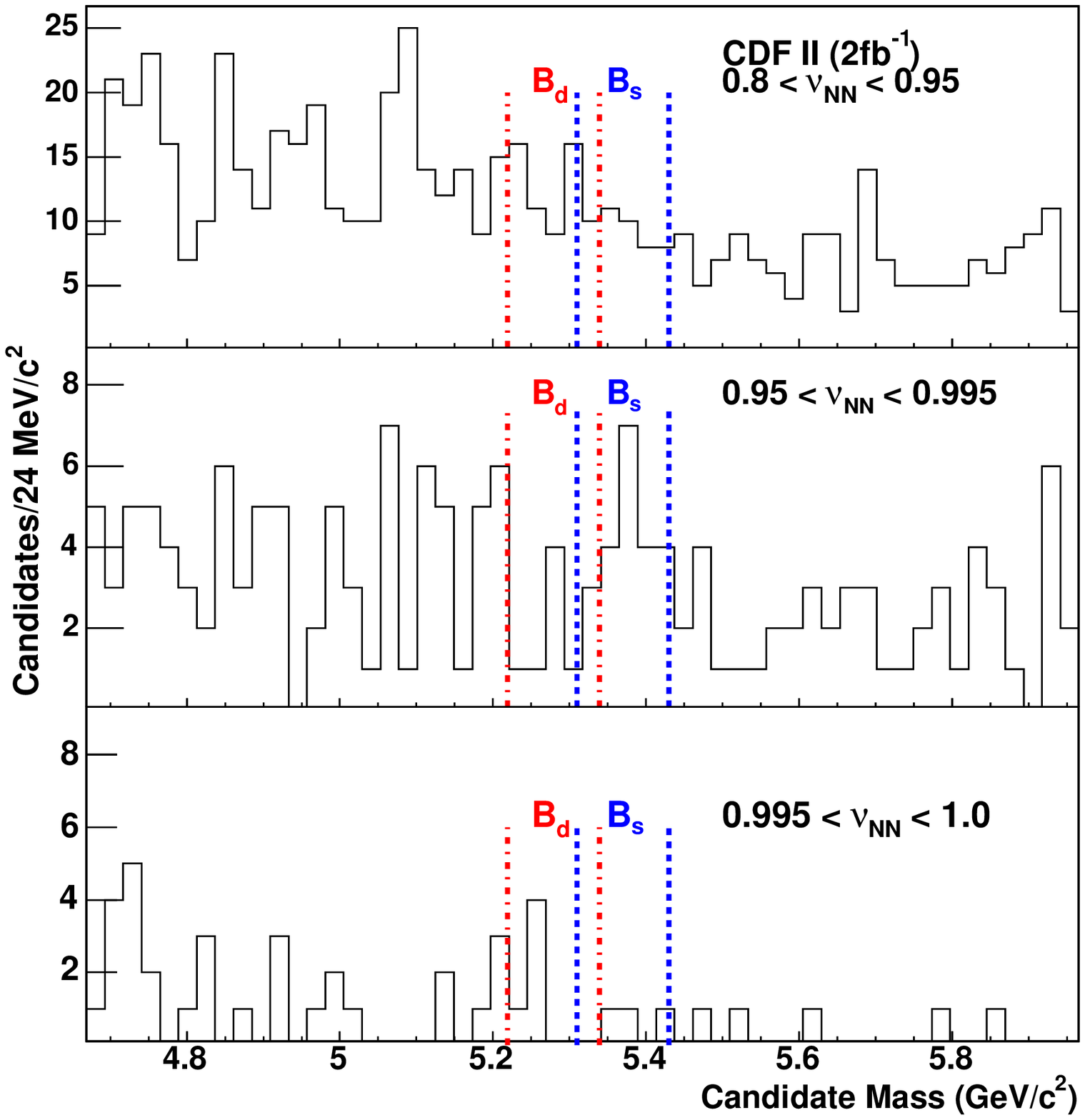}
}
\end{center}
\caption{Invariant mass spectrum of \Bsmumu candidates
measured by D0 in two run ranges (left) and by CDF in three bins of neural network output (right).}
\label{fig:BsMuMu}
\end{figure}

Both Tevatron experiments optimize the selection of \Bsmumu candidates using simulated signal
events and background events from mass sidebands.
While D0 combines the discriminant variables in a likelihood ratio, CDF uses a neural network (NN).
It was checked on background samples that the NN does not introduce a selection bias.
Both experiments estimate the combinatorial background by a fit to the mass sidebands.
The contribution from decays of $B$ mesons to two light hadrons, which could peak in the signal
mass region, was estimated to be an order of magnitude lower than the combinatorial background.
To obtain an absolute branching ratio the number of signal events is normalized to the
high-statistics $B^+ \rightarrow J/\psi K^+$ mode.
For the limit calculation CDF splits the data sample in three bins in NN output and five bins in mass
which improves the sensitivity by 15\% compared to using just one bin.
Both experiments do not see a significant excess (Fig.\ \ref{fig:BsMuMu}).
The 90\% confidence level (CL) limits calculated with a Bayesian method for a data sample of 2 fb$^{-1}$ per
experiment are $\mathcal{B}(\Bsmumu) < 7.5 \times 10^{-8}$ (D0)\cite{D0:Bsmumu} and 
$\mathcal{B}(\Bsmumu) < 4.7 \times 10^{-8}$ (CDF)\cite{Aaltonen:2007kv}.
Because of the good mass resolution of the tracking system CDF is able to 
separate $B^0_s$ and $B^0$ mesons and to quote
a 90\% CL limit on the $B^0$ decay of $\mathcal{B}(\Bdmumu) < 1.5 \times 10^{-8}$.

\subsection{\D0mumu}

The SM box and penguin processes of the \D0mumu decay are much more suppressed by the GIM mechanism
than in the \Bsmumu case.
Therefore long distance processes like the decay via hadronic resonances and photons are dominant
resulting in a predicted branching ratio\cite{Burdman:2001tf} of $\mathcal{B}(\D0mumu) \ge 4 \times 10^{-13}$.
While no significant enhancement is expected in $R$-parity conserving SUSY models,
branching ratios up to $10^{-6}$ are possible if $R$-parity is violated.

\begin{figure}[h]
\begin{minipage}{0.32\textwidth}
\includegraphics[width=0.99\textwidth]{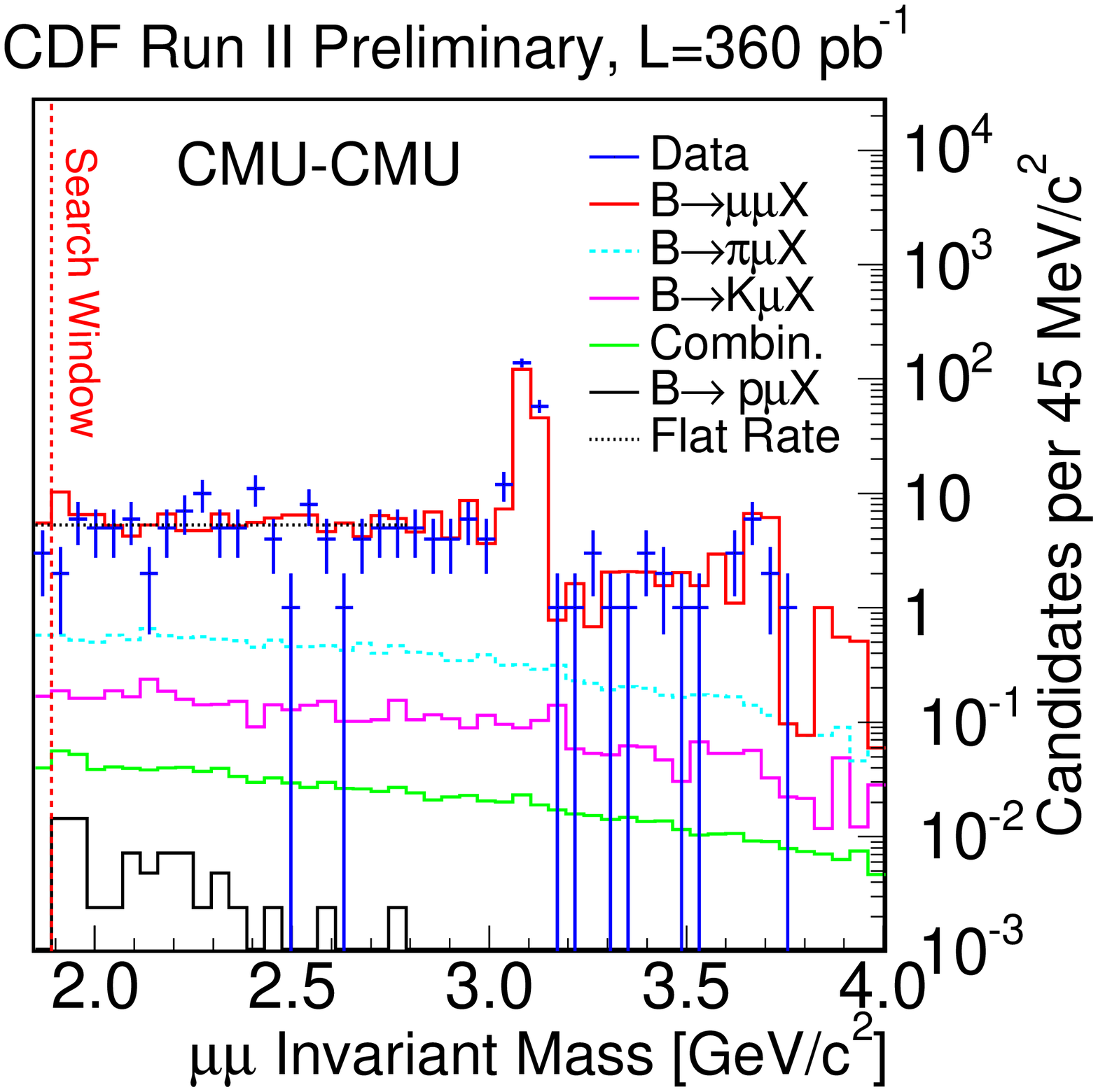}
\caption{Invariant mass spectrum of \D0mumu candidates in the CMU-CMU acceptance region.}
\label{fig:D0MuMu}
\end{minipage}
\hspace*{0.015\textwidth}
\begin{minipage}{0.66\textwidth}
\begin{center}
\includegraphics[width=0.49\textwidth]{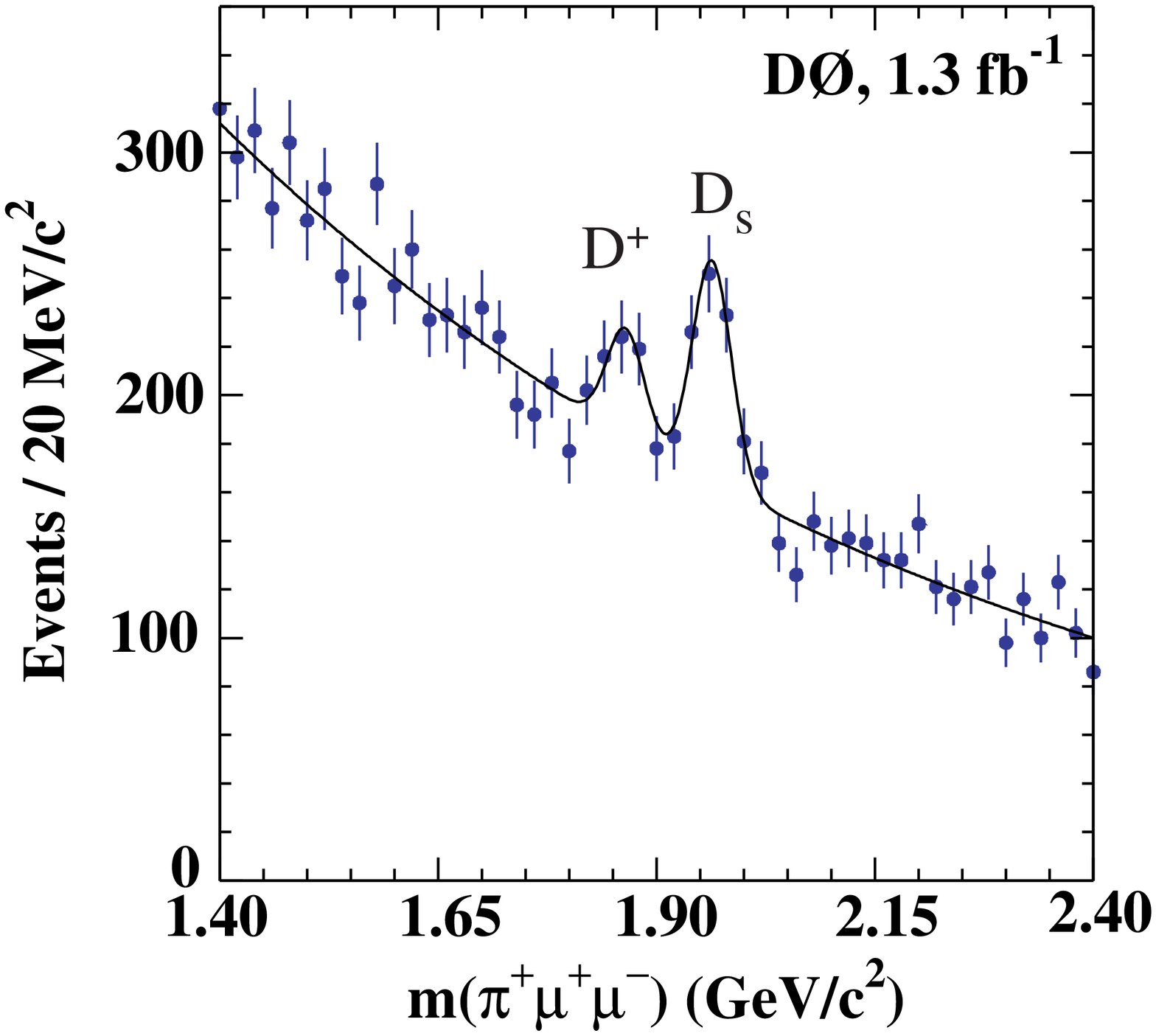}
\includegraphics[width=0.49\textwidth]{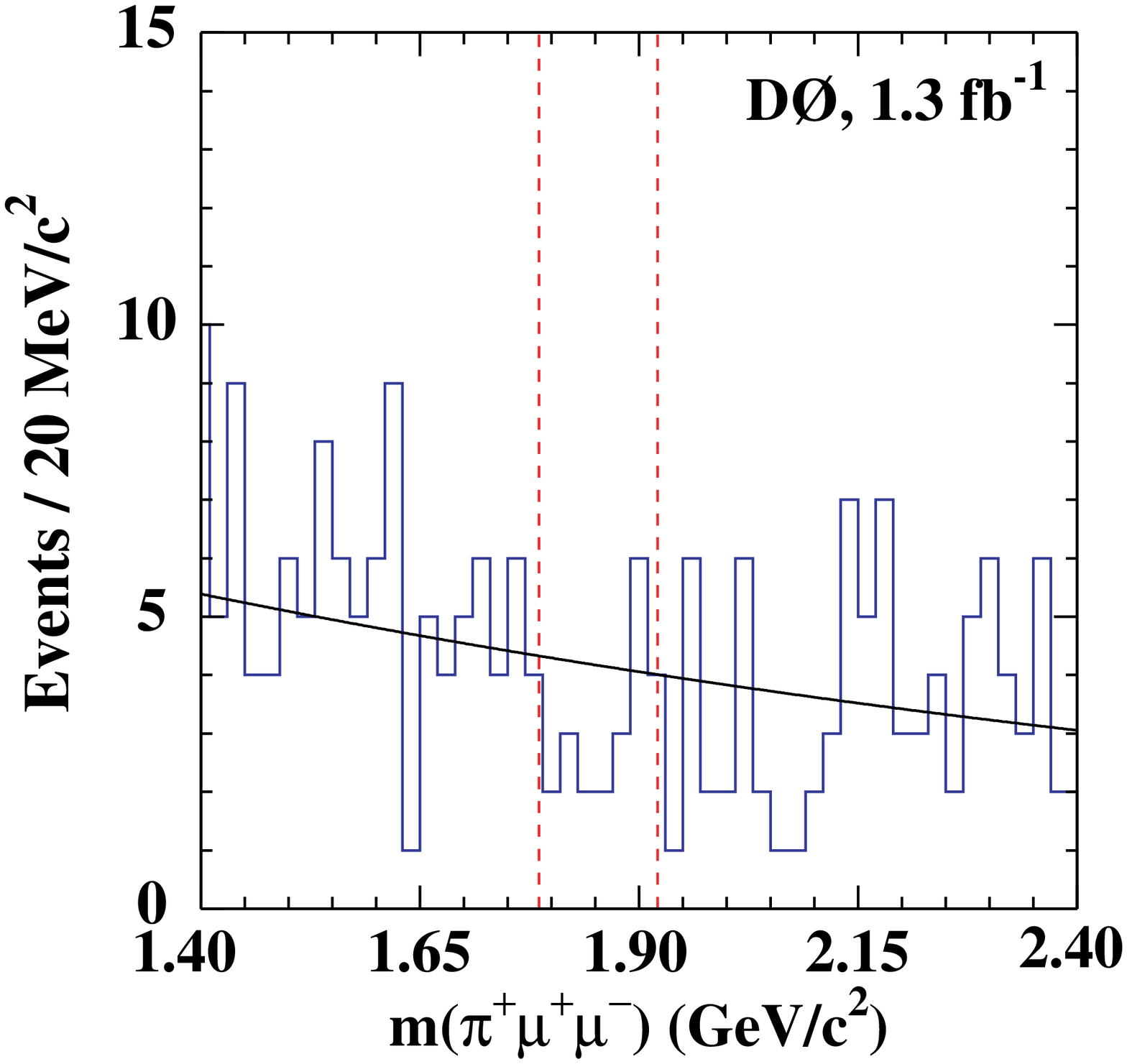}
\end{center}
\caption{Invariant $\pi^+\mu^+\mu^-$ mass spectrum of for $\mu^+\mu^-$ candidates
in the $\phi$ mass region (left) and outside the $\phi$ mass region (right).}
\label{fig:DPiMuMu}
\end{minipage}

\end{figure}

CDF selects candidates of \D0mumu decays by a trigger on displaced tracks which allows to use the
$D^0 \rightarrow \pi^+\pi^-$ mode for normalization.
Muons are identified using muon chambers in the pseudorapidity ranges $|\eta|<0.6$ (CMU) and
$0.6<|\eta|<1$ (CMX).
Background events are reduced by requiring the $D^0$ to come from a $D^*$ decay and by cutting
on a lifetime information based probability ratio between signal events and $B \rightarrow \mu^+\mu^-X$ events,
the dominant background.
In a data sample of 360 pb$^{-1}$ the observed numbers of events of 3, 0 and 1 in the acceptance
regions CMU-CMU, CMU-CMX and CMX-CMX are consistent with the background expectations of $4.9 \pm 1.5$,
$2.7 \pm 1.0$ and $1.0 \pm 0.5$, respectively (Fig.\ \ref{fig:D0MuMu}).
The obtained 90\% CL Bayesian limit\cite{CDF:D0mumu} is $\mathcal{B}(\D0mumu) < 4.3 \times 10^{-7}$.

\subsection{\Dpimumu}

The \Dpimumu decay is, like the \D0mumu decay, dominated by long distance processes.
But the size of this contribution depends on the dimuon mass.
The selection of non-resonant dimuon masses therefore increases the sensitivity to new physics contributions,
in particular from $R$-parity violating SUSY.

In the first part of the analysis D0 selects events with $m(\mu^+\mu^-)$ in the $\phi$ meson mass region to
establish a resonant decay signal.
In the $m(\pi^+\mu^+\mu^-)$ spectrum a $D^+_s$ signal and a $D^+$ signal are observed (Fig.\ \ref{fig:DPiMuMu})
with a statistical significance  of 8$\sigma$ for both combined and 4.1$\sigma$ for the $D^+$ alone.
The measured value\cite{Abazov:2007kg} of  
$\mathcal{B}(D^{+} \rightarrow \phi\pi^+ \rightarrow \mu^+\mu^-\pi^+) = (1.8 \pm 0.5 \pm 0.6) \times 10^{-6}$,
using the resonant $D^+_s$ decay as normalization, is in good agreement with\cite{Yao:2006px} 
$\mathcal{B}(D^{+} \rightarrow \phi\pi^+) \cdot \mathcal{B}(\phi \rightarrow \mu^+\mu^-) = (1.86 \pm 0.26) \times 10^{-6}$.

For the search for the non-resonant decay events with $m(\mu^+\mu^-)$ in the $\phi$ mass region are excluded.
The observed number of 19 events in the $m(\pi^+\mu^+\mu^-)$ search window is consistent with the
background expectation of $25.8 \pm 4.6$ events (Fig.\ \ref{fig:DPiMuMu}).
By normalizing to the resonant $D^+$ decay a 90\% CL Bayesian limit of 
$\mathcal{B}(\Dppimumu) < 3.9 \times 10^{-6}$ is determined from a data sample of 1.3 fb$^{-1}$.

\subsection{\Lbppi and \LbpK}

The decays \Lbppi and \LbpK are allowed at tree level in the SM, but are suppressed by the small value
of the involved CKM matrix element $V_{ub}$.
Therefore loop diagram processes can contribute at a magnitude that is comparable to the tree diagram process.
The interference of these amplitudes can lead to a sizeable direct CP violation.
In the SM an $A_{CP}$ value of $\mathcal{O}(10\%)$ is predicted.
While $R$-parity violating SUSY processes would enhance the branching ratio from $\mathcal{O}(10^{-6})$
up to $\mathcal{O}(10^{-4})$ they would at the same time reduce $A_{CP}$ by one order of magnitude.

\begin{figure}[h]
\begin{center}
\includegraphics[width=0.35\textwidth]{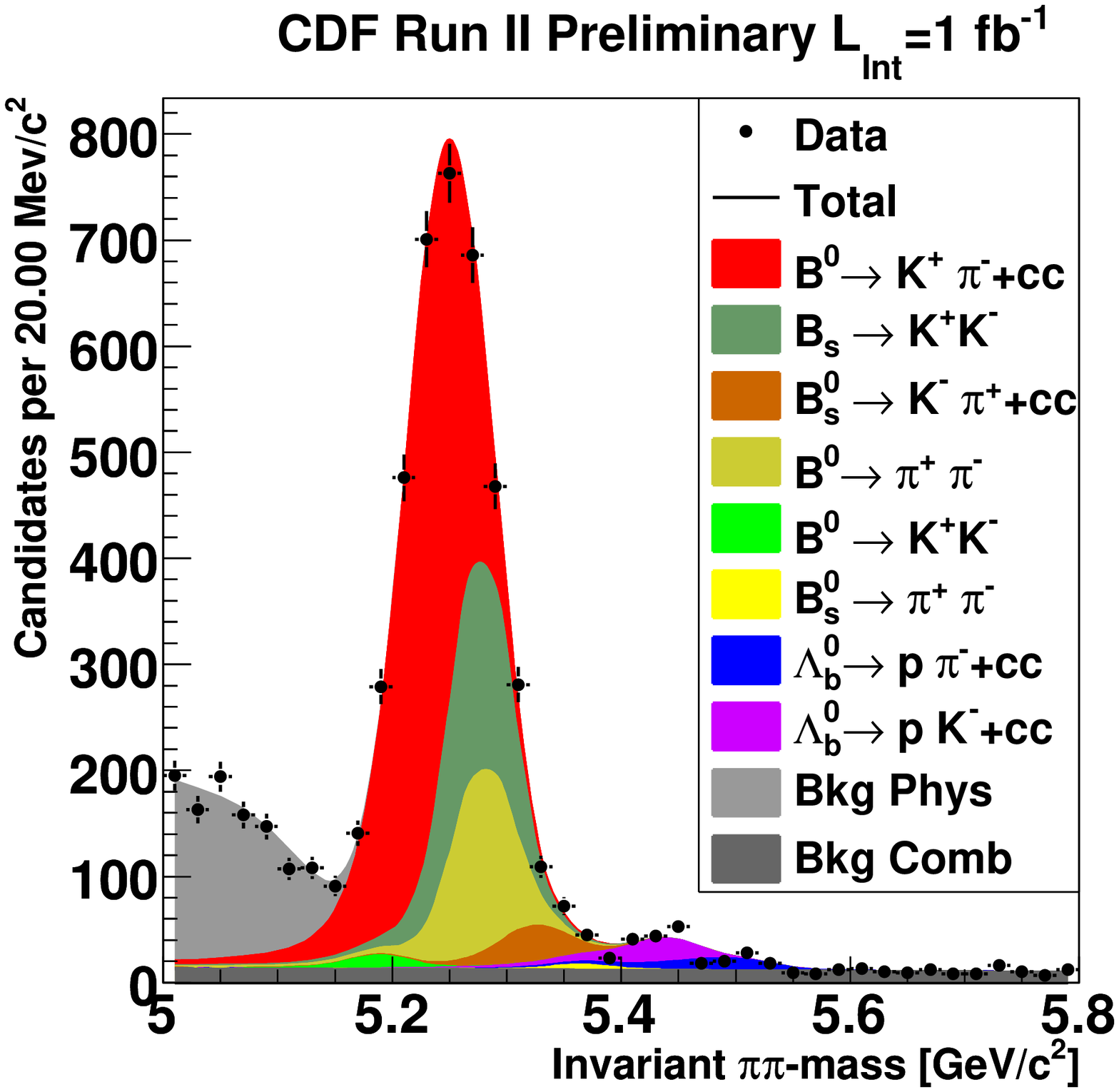}
\hspace*{0.1\textwidth}
\includegraphics[width=0.35\textwidth]{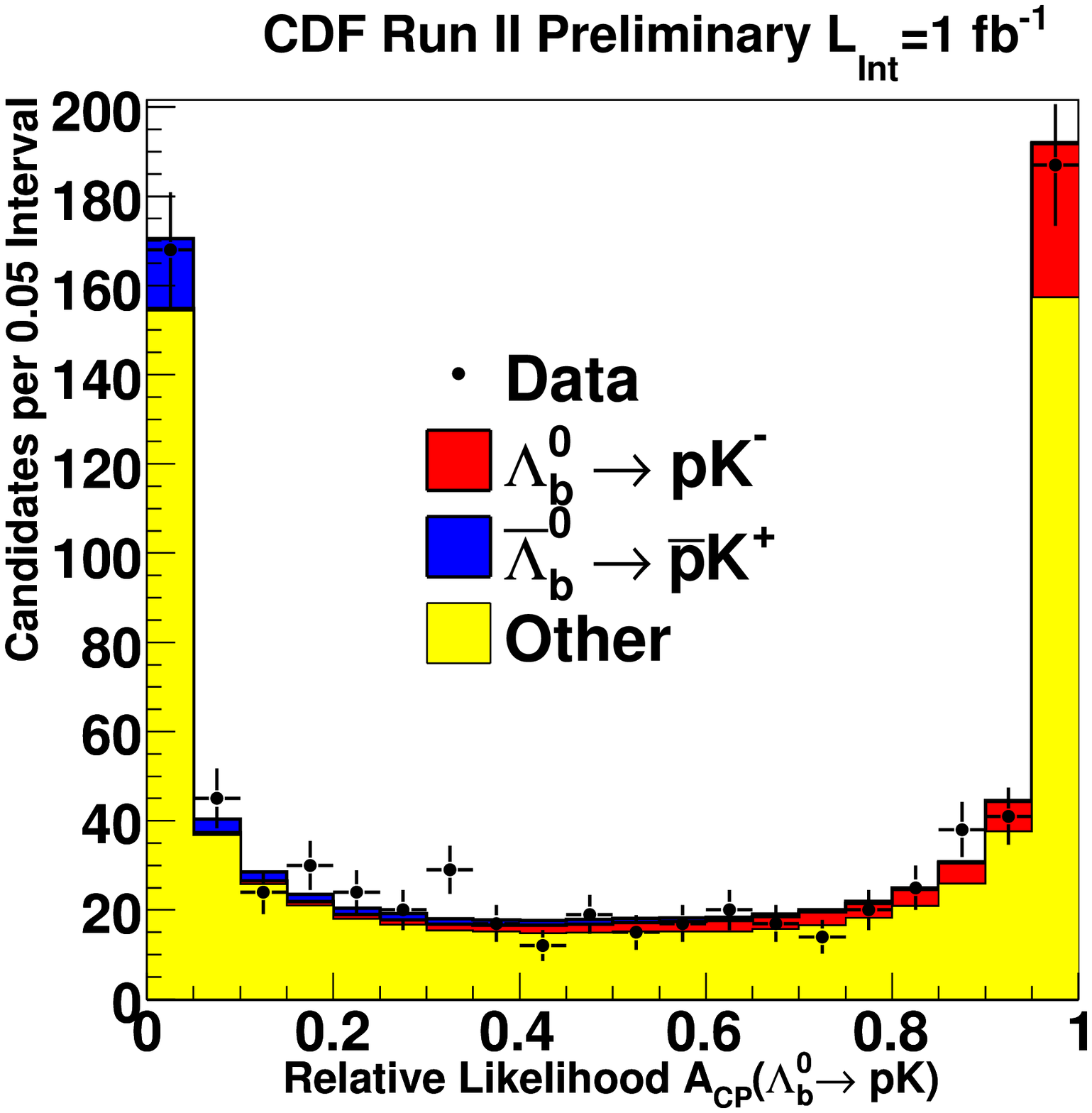}
\end{center}
\caption{Invariant mass spectrum for $\pi^+\pi^-$ mass assignment (left)
and relative probability density function (pdf) of \LbpK:
$\mbox{pdf}(\Lambda_b^0) / [\mbox{pdf}(\Lambda_b^0)+\mbox{pdf}(\bar{\Lambda}_b^0)]$ (right).}
\label{fig:Lbph}
\end{figure}

Both rare $\Lambda_b$ decays were first observed in a CDF analysis of rare $B$ meson decays
to two light hadrons\cite{CDF:Bhh}.
The analysis technique developed there is reused here.
It involves an unbinned likelihood fit of invariant mass under dipion hypothesis (Fig.\ \ref{fig:Lbph} left),
relations between daughter particle momenta, and particle identification information provided by
the specific ionization energy loss in the tracker.
With the decay $B^0 \rightarrow K^+\pi^-$ as normalization the quantities
$\mathcal{B}(\Lbppi)/\mathcal{B}(B^0 \rightarrow K^+\pi^-) \cdot f_{\Lambda_b^0}/f_{B^0} = 0.0415 \pm 0.0074 \pm 0.0058$ and
$\mathcal{B}(\LbpK)/\mathcal{B}(B^0 \rightarrow K^+\pi^-) \cdot f_{\Lambda_b^0}/f_{B^0} = 0.0663 \pm 0.0089 \pm 0.0084$
are measured\cite{CDF:Lbph} in a data sample of 1 fb$^{-1}$ 
where $f_{\Lambda_b^0}/f_{B^0}$ is the $\Lambda_b^0$ to $B^0$ production ratio.
Taking the production ratio measured by CDF\cite{Aaltonen:2008zd} and the known $B^0 \rightarrow K^+\pi^-$ 
branching ratio\cite{Yao:2006px} one obtains absolute branching ratios
of $\mathcal{B}(\Lbppi) = (1.4 \pm 0.3 ^{+0.9}_{-0.5}) \times 10^{-6}$ and 
$\mathcal{B}(\LbpK) = (2.2 \pm 0.3 ^{+1.4}_{-0.8}) \times 10^{-6}$ which are in good agreement
with the SM predictions of $1 \times 10^{-6}$ and $2 \times 10^{-6}$, respectively\cite{Mohanta:2000nk}.

To measure the CP asymmetry $A_{CP}(\Lambda_b^0 \rightarrow p\,h^-, h = \pi\mbox{ or }K) =
[\mathcal{B}(\Lambda_b^0 \rightarrow p\,h^-) - \mathcal{B}(\bar{\Lambda}_b^0 \rightarrow \bar{p}\,h^+)] /
[\mathcal{B}(\Lambda_b^0 \rightarrow p\,h^-) + \mathcal{B}(\bar{\Lambda}_b^0 \rightarrow \bar{p}\,h^+)]$
the relative efficiencies are determined from inclusive $\Lambda^0 \rightarrow p\pi^-$ and 
$\bar{\Lambda}^0 \rightarrow \bar{p}\pi^+$ decays.
While the result of $A_{CP}(\Lambda_b^0 \rightarrow p\pi^-) = 0.03 \pm 0.17 \pm 0.05$ is well consistent
with no CP asymmetry, the value of $A_{CP}(\Lambda_b^0 \rightarrow pK^-) = 0.37 \pm 0.17 \pm 0.03$
is about 2$\sigma$ away from zero.
Fig.\ \ref{fig:Lbph} illustrates the asymmetry as well as the good description of the data by the fit
and the powerful $\Lambda_b^0$/$\bar{\Lambda}_b^0$ separation.

\section{Conclusions}
New world's best limits on the branching ratio of the rare decays \Bsmumu, \Bdmumu, \D0mumu and 
\Dppimumu were presented.
Furthermore the first branching ratio and CP asymmetry measurement of charmless hadronic $\Lambda_b$
decays were shown.
These Tevatron results can impose stringent constraints on physics beyond the SM.
A further significant reduction of the new physics models parameter space can be expected
as more data is taken and analyzed.

\end{document}